# Surface Thermal Gradients Activated by Enhanced Molecular Absorption in Mid-infrared Vertical Antenna Arrays

Andrea Mancini[1], Valeria Giliberti[2], Alessandro Alabastri[3], Eugenio Calandrini[4], Francesco De Angelis[4], Denis Garoli[4*] and Michele Ortolani[1,2*]

[1] Dipartimento di Fisica, Sapienza University of Rome, Piazzale Aldo Moro 5, 00185 Rome, Italy

[2] Center for Life NanoSciences, Istituto Italiano di Tecnologia (IIT), Viale Regine Elena 291, 00185 Rome, Italy

[3] Department of Electrical and Computer Engineering, Rice University, 6100 Main Street MS 366, Houston, TX 77005, USA

[4] Plasmon Nanotechnologies Department, Istituto Italiano di Tecnologia (IIT), 16163 Genoa, Italy

*corresponding author e-mails: denis.garoli@iit.it , michele.ortolani@roma1.infn.it

***Abstract.*** *We investigate local heat generation by molecules at the apex of polymer-embedded vertical antennas excited at resonant mid-infrared wavelengths, exploiting the surface enhanced infrared absorption (SEIRA) effect. The embedding of vertical nanoantennas in a non-absorbing polymer creates thermal isolation between the apical hotspot, the locus of heat generation, and the heat sink represented by the substrate. Vibrational mid-infrared absorption by strongly absorbing molecules located at the antenna apex then generates nanoscale temperature gradients at the surface. We imaged the thermal gradients by using a nano-photothermal expansion microscope, and we found values up to 10 K/μm in conditions where the radiation wavelength resonates with both the molecule vibrations and the plasmonic mode of the antennas. Values up to 1000 K/μm can be foreseen at maximum quantum cascade laser power. The presented system provides a promising thermoplasmonic platform for antenna-assisted thermophoresis and resonant mid-infrared photocatalysis.*

Generation of heat and control of temperature profiles at the nanoscale have attracted considerable interest in the last years[1-4]. Remote temperature control by focused light beams allows for fast and contact-less generation of heat at specific locations, and therefore it is currently under study for applications such as photocatalysis[5-7], microscale thermophoresis[2,8-10], induction of phase transitions in nanoscale materials[11,12], and medical therapy[13,14]. All-dielectric photothermal effects can hardly go beyond the diffraction limit[15], therefore thermoplasmonics, i.e. the generation of heat from metal nanoparticles and nanoantennas[1,3-7], has recently arisen as the technique of choice for nanoscale photothermal heat generation with significant efficiency in small volumes[3,4]. In some advanced applications, however, the nanoscale control of thermal gradient geometry and intensity is of key importance[16] therefore periodic optical antenna arrays have been proposed for long-range ordered heat generation pattern[17].

Most thermoplasmonic devices are based on strong dissipation of oscillating electric currents in the metal (plasmon dissipation)[1], which represents the optical equivalent of the Joule heating effect at dc. In the visible range, gold nanoparticles dispersed in liquids or dielectric matrixes can produce nanoscale hotspots[3,13,14,16]. In the mid-infrared (mid-IR) range, periodic antenna arrays fabricated by high-resolution lithography on solid substrates in both planar[18,19] and vertical versions[4,20-23] have been developed mostly for molecular sensors based on the Surface Enhanced Infrared Absorption[24,25] (SEIRA) mechanism. Periodic arrays provide the opportunity to explore new designs for heat generation, in which the coherent enhancement of the electromagnetic (e.m.) field due to constructive interference at specific wavelengths can be exploited to produce strong temperature gradients[16,17,23]. However the generation of nanoscale thermal gradients by lithographic metal antennas anchored to solid substrates leads to a contradiction: to obtain a strong gradient, the light-absorbing material should feature low thermal conductivity, but metal surfaces and solid substrates generally feature high thermal conductivity and diffusivity and, as such, they tend to homogenize the temperature profiles. In this paper, we present a device[23] that realizes sub-micrometer thermal gradients through SEIRA heating, solving the above mentioned contradiction. The fundamental mechanism of SEIRA heating is the non-radiative decay of highly excited vibrational states of the molecule: not only it physically differs from plasmon dissipation, but it also takes place outside the crystal lattice the metal. Therefore, the nanoscale thermal conductivities and diffusivities of the molecules prevail on those of the metal and substrate. With our device, we have obtained an order of magnitude increase in both intensity and abruptness of surface thermal gradients if compared to pure plasmon dissipation in the same plasmonic structures.

The plasmonic structures studied in this work consist of square arrays of vertical gold-coated rods placed on a gold-coated substrate, 2.2 μm tall, 360 nm in diameter, with different array pitch $P$ of 3.0, 3.5 and 4.0 μm. A top-view electron microscopy image of one of the arrays is reported in Fig. 1a (see Refs. 19-21 for the fabrication process). When illuminated with mid-infrared radiation at specific resonant frequencies, the structures generate field-enhancement hotspots at the antenna apexes[20-23]. Embedding the structures in a polymer layer with low thermal conductivity and weak IR absorption (AZ 5214E by MicroChemicals GmbH, thickness 2.1 μm) results in a quasi-planar surface (Fig. 1b). To generate heat by SEIRA, we deposited a further poly-methylmethacrylate (PMMA) cap layer (thickness 0.1 μm) that spatially overlaps with e.m. hotspots at the antenna apexes (see Fig. 1c). PMMA has a strong vibrational absorption at $\omega_{res} = 1730$ cm$^{-1}$ matching the coherent plasmonic resonance frequency of the array with $P = 3.5$ μm[21,23]. The three-dimensional temperature distribution is calculated by joint e.m. and thermal simulation[4,23] (COMSOL multiphysics, see example in Fig. 1c). The two-dimensional surface temperature profile is experimentally imaged (see example in Fig. 1d) with a nano-photothermal expansion microscope (AFM-IR technique[26-30], NanoIR2 by Anasys Instruments) based on an atomic force microscope (AFM) and a tunable mid-IR quantum cascade laser (QCL, MIRcat by Daylight Solutions). The focused QCL impinging on the sample with an incidence angle of 70° to the surface normal illuminates the antenna array underneath the AFM tip, with electric field polarization almost parallel to the antenna axis. While gold-coated tips are often employed in AFM-IR setups when studying non-resonant samples[31] so as to exploit the lightning rod effect at the metal tip apex, here uncoated silicon probe tips have been employed in order to leave unperturbed the field enhancement provided by the sample itself[30]. The amplitude of the AFM cantilever oscillations monitors the thermal expansion of the molecules triggered by the local IR absorption[26-28]. In the case of chemically homogeneous IR-absorbing films (here, the spin-coated PMMA layer) any contrast in the AFM-IR absorption maps is produced by temperature gradients within the sample[32]. In the AFM-IR image acquired at $\omega_{res} = 1730$ cm$^{-1}$ (Fig. 1d), SEIRA heating by the PMMA molecules located at the antenna hotspots results in a square array of sub-micrometric thermal hotspots at the surface, not directly related to topography features (compare Fig. 1d with Fig. 1b).

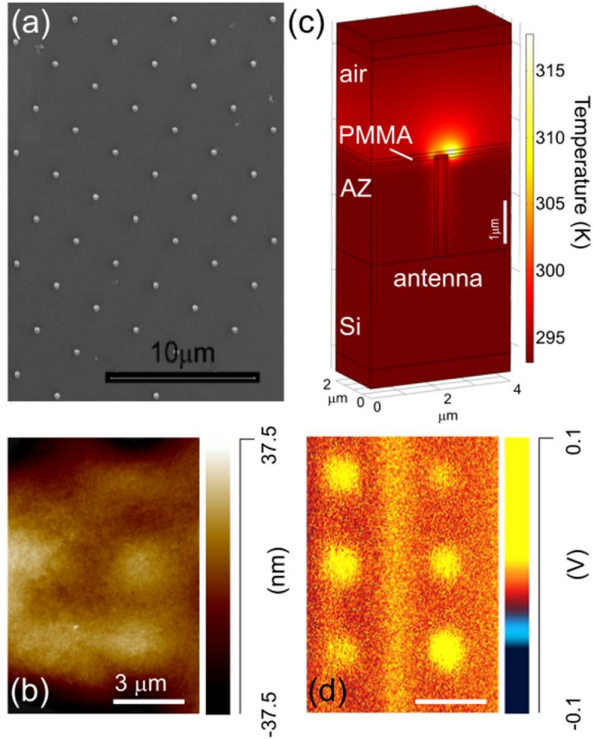

**FIG 1. (a)** Electron microscopy image of the vertical antenna array with $P$ = 3.5 μm seen from top before embedding it into polymers. **(b)** AFM topography after the polymer-embedding procedure. **(c)** Simulation of the temperature increase obtained by focused mid-IR laser illumination at the plasmonic resonance frequency of the array, matching the PMMA vibration frequency $\omega_{res}$ = 1730 cm$^{-1}$. A single unit cell of the polymer-embedded array is shown. The incidence angle is set to 70° to the surface normal and the illumination direction is from the left. **(d)** AFM-IR map of the same area as in (b) with the laser tuned at $\omega_{res}$.

Extensive e.m. and thermal simulations as a function of $\omega$ and $P$ with artificially varied molecule absorption strength confirm that PMMA molecules are responsible for the surface temperature patterns seen in our experiments (Fig. 1d). AZ molecules located along the antenna shaft, where additional field enhancement regions exist at $\omega_{res}$ = 1730 cm$^{-1}$, are found to produce negligible SEIRA heating because AZ is a weak mid-IR absorber (see Fig. 1c). Also, although the total optical power absorbed by the PMMA molecules in the hotspot and by the metal is comparable[23], plasmon dissipation does not significantly contribute to the surface thermal gradients, due to the high thermal conductivity and diffusivity of gold and to the heat-sink effect of the solid substrate (silicon). The substrate heat-sink effect severely limits the thermal gradients that can be obtained in more conventional planar antenna arrays, as previously observed by AFM-IR studies[28-30,32].

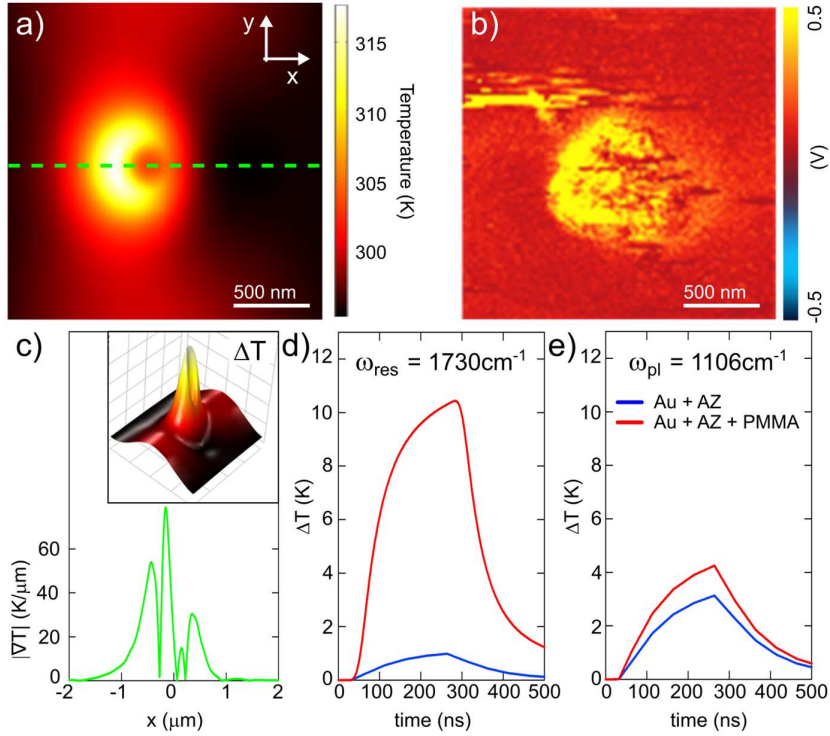

**FIG 2.** (a) Simulated surface temperature map and (b) experimental nano-photothermal expansion map measured at $\omega_{res} = 1730$ cm$^{-1}$. (c) Temperature gradient calculated from data in (a) along the green dashed line. Inset of (c): three-dimensional version of the color plot in (a) highlighting the gradient intensity. (d-e) Simulated temperature increase above the antenna rim (x =0.18 μm) at $\omega_{res} = 1730$ cm$^{-1}$ and at $\omega_{pl} = 1106$ cm$^{-1}$. Red curves: actual experimental structure with PMMA layer; blue curves: same but without PMMA layer, evidencing the small contribution due to plasmon dissipation at $\omega_{res} = 1730$ cm$^{-1}$ and the small contribution due to non-absorbing PMMA at $\omega_{pl} = 1106$ cm$^{-1}$.

In Fig. 2 we show details of the simulations of the temperature profile (Fig. 2a) and the corresponding experimental map (Fig. 2b). Our numerical simulations perfectly reproduce the asymmetric shape of the thermal hotspot, which is stronger on the side opposite to the quasi-grazing incidence illumination direction ($x < 0$) due to stronger field enhancement on that side of the antenna. The small circular spot at lower temperature at the antenna center ($x = y = 0$) is due to heat conduction through the antenna shaft. Outside the antenna rim, the temperature drops as the inverse of the distance from the hotspot[1] and produces sharp surface gradients. The gradient can be explicitly computed from the simulation data of Fig. 2a as:

$$|\nabla T| = \sqrt{\left(\frac{\partial T}{\partial x}\right)^2 + \left(\frac{\partial T}{\partial y}\right)^2} \quad (1).$$

Along the $y = 0$ line (green dashed line in Fig. 2a), the result of Eq. 1 reported in Fig. 2c is independent on variations along $y$ by symmetry, and maxima are obtained at three locations: $x = -0.18$ μm (above the antenna rim) and $x = \pm 0.4$ μm (approximate limits of the field enhancement region). The gradient is clearly nonzero in an approximately circular contour of 0.7 μm radius. The value of the gradient maxima of the order of 10 K/μm for the laser power of 20 mW used here. Projecting our results towards practical applications, gradient intensities of the gradient as high as 1000 K/μm could likely be obtained with maximum QCL power of a few Watts. In the present experiment we limit the QCL power to 20 mW to avoid saturation of the mechanical oscillation amplitude of the AFM cantilever and, eventually, sample melting, which could instead be pursued in purposely engineered thermoplasmonic reaction cells[6-7].

The red curves in Fig. 2d and 2e report the simulated temperature dynamics during the 260 ns long QCL pulse above the antenna rim at $\omega_{res} = 1730$ cm$^{-1}$ and at a different plasmonic resonance of the array $\omega_{pl} = 1106$ cm$^{-1}$ where PMMA does not strongly absorb. The simulations are repeated for an identical configuration without the absorbing PMMA top layer (blue curves) in order to demonstrate that the temperature increase is ten times stronger when PMMA vibrations are excited, i.e. that SEIRA heating dominates over Joule heating at $\omega_{res} = 1730$ cm$^{-1}$. The timescale of temperature increase obtained from the dynamical simulations is of the order of 100 ns, while the laser pulse length was set to 260 ns as in the experiment[23]. This means that longer pulse lengths or continuous wave illumination would increase temperature scales even further if compared to the present experiment, at equal laser power[27]. To summarize, in presence of absorbing molecules in the apical antenna hotspot SEIRA heating can produce strong surface thermal gradients that could linearly scale with laser excitation power.

We further study the spectral response of the embedded nanoantenna device by positioning the AFM tip on the highest temperature location and scanning the QCL emission frequency over its tuning range 1100-1900 cm$^{-1}$ and detuning the plasmonic resonance from the PMMA vibration by analyzing arrays with different *P* by AFM-IR spectroscopy (Figure 3a-c). IR reflection spectroscopy[23] has shown that the plasmonic mode frequency, which is not directly detected by AFM-IR because plasmon dissipation does not significantly contribute to temperature increase, shifts with *P* due to its "spoof" surface plasmon nature[22]. The strongest enhancement of the PMMA vibration at 1730 cm$^{-1}$ is observed in Fig. 3b, because the array with *P* = 3.5 μm indeed has the plasmonic resonance very close to 1730 cm$^{-1}$. In Figure 3a and 3c the absorption enhancement is reduced because of detuning. For comparison, in Fig. 3d (spectrum taken in a region of the chip far from the antenna arrays) the three main PMMA vibrational fingerprints at 1200, 1450 and 1730 cm$^{-1}$ are equally visible due to the lack of any plasmonic field enhancement.

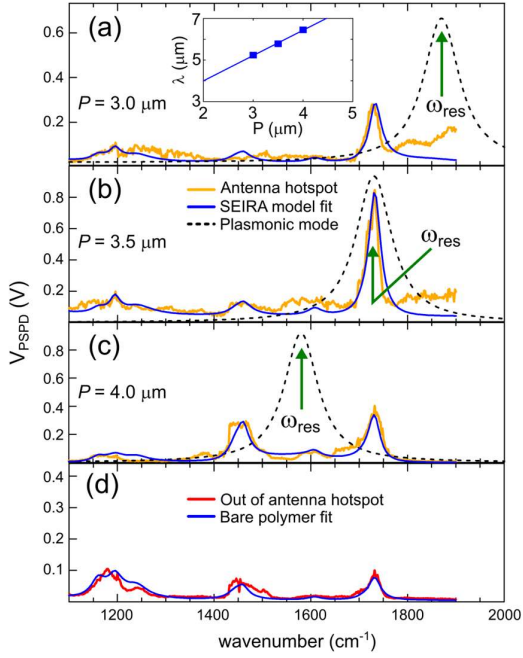

**FIG 3.** AFM-IR spectra measured with the tip at the highest temperature location above the antenna rim of one antenna for the three arrays (a-c) and at a location far from the arrays (d). Blue curves: best-fit of Eq. 2 to the data. The SEIRA intensity enhancement lineshapes extracted from the best-fits are reported as black dashed curves. Inset of panel (a): plasmon resonance wavelength displaying a linear dependence on $P$. All experimental spectra are normalized for the frequency-dependent QCL power.

A model with two interplaying terms can snatch the essential of the spectral absorbance lineshape $A(\omega)$ resulting from SEIRA heating: direct absorption from all molecules in the illuminated region according to the Beer-Lambert's law (product of the absorption coefficient $\alpha_i$ times the polymer layer thickness $d_i$) and plasmon-enhanced absorption in the antenna hotspots:

$$A(\omega) = C_0 + C_1 \cdot \sum_i [d_i \alpha_i(\omega) + c_i \alpha_i(\omega) L(\omega) r_{hs}] \qquad (2)$$

where $L(\omega)$ is the intensity enhancement spectrum defined by the spectral response of the plasmonic mode; $r_{hs}$ is a characteristic linear dimension of the hotspot; $c_i$ is the relative volume concentration of a given molecular species in the hotspot; the index $i$ runs over all substances present in the hotspot (here, the two polymers PMMA and AZ). The SEIRA enhancement lineshape $L(\omega)$ is assumed to be Lorentzian[25], with fixed peak height of 14 and width of 70 cm$^{-1}$ as determined in Ref. 23. Obviously, $L(\omega)$ was set to zero in the fit of Fig. 3d. Separate IR transmission spectroscopy experiments on spin-coated films from the same source provided the absorption coefficient spectra $\alpha_{AZ}(\omega)$, $\alpha_{PMMA}(\omega)$. Finally, we introduce two calibration coefficients $C_0$ and $C_1$, to fit the model to the AFM-IR data expressed as the AFM position sensitive photodetector voltage component $V_{PSPD}$ at the mechanical resonance frequency of the cantilever[28]. Fig. 3 discloses the results of our analysis and demonstrates the linear e.m. scaling and frequency tunability of the proposed approach: blue curves represent the best fit of Eq. 2 to our data with $c_{PMMA} \sim 0.9$, $c_{AZ} \sim 0.1$, $r_{hs} \sim 50$ nm, $d_{PMMA} \sim 100$nm, $d_{AZ} \sim 2100$ nm. The plasmon resonance frequency is then left as the only relevant free parameter and the resulting resonance wavelength $\lambda_{res} = 2\pi/\omega_{res}$ is plotted in the inset of Fig. 3a as a function of $P$, demonstrating linear e.m. scaling and frequency tunability of the proposed approach.

As possible applications, we propose that polymer-embedded vertical antenna arrays may be used as antenna-assisted thermoplasmonic surfaces for microscale thermophoresis[9,10] on the relatively flat PMMA surface. Alternatively, the PMMA layer could be substituted with active chemical species, possibly in liquid form, and the AZ-embedded array could be employed for tunable mid-IR photocatalysis, where only those molecules with vibrational fingerprints resonant with the plasmonic mode frequency are substantially heated up through SEIRA. This would provide chemical selectivity to thermoplasmonic reaction cells, an asset not available to existing broadband mid-IR perfect absorbers[28], carbon black or nanoporous gold[22]. Finally, vertical antennas can punch through living cell membranes[34] and used to heat up selected molecular species in the cytoplasm by SEIRA rather than by non-selective plasmon dissipation[35]

In conclusion, we studied by nanoscale photothermal microscopy the temperature gradients generated under mid-infrared laser illumination at the surface of vertical nanoantenna arrays embedded in polymer layers. The vertical geometry of the sample allows the displacement of the thermal hotspots away from the substrate, effectively canceling the high thermal conductivity and diffusivity of the metallic antenna and of the solid substrate. The relevant heat is generated directly in the polymer molecules by SEIRA, and not by plasmon dissipation in the metal, leading to observed gradients of 10 K/µm and perspective values up to 1000 K/µm. Additionally, the presented photothermal transducers feature chemical selectivity, provided by SEIRA through the mid-IR vibrational fingerprints.